\documentclass[12pt,preprint]{elsarticle}

\usepackage{amsmath,amssymb}
\usepackage{xspace}
\usepackage{psfrag,graphicx}
\usepackage{lineno}
\makeatletter
\g@addto@macro\endfrontmatter{\enlargethispage{-2\baselineskip}}
\makeatother

\journal{Astroparticle Physics}

\newcommand{\meanlnA}{\ensuremath{\langle\ln A\rangle}\xspace}
\newcommand{\xmaxs}{\ensuremath{X_\mathrm{max}}\xspace}
\newcommand{\loge}{\ensuremath{\log(E/\mathrm{eV})}\xspace}
\newcommand{\gsm}{\ensuremath{\mathrm{g\,cm^{-2}}}\xspace}
\newcommand{\iah}[1]{\ensuremath{I^A_\text{h}(#1)}}
\newcommand{\logenr}[2]{\ensuremath{\log(E/\mathrm{eV})=#1-#2}}
\newcommand{\logen}[1]{\ensuremath{\log(E/\mathrm{eV})=#1}}
\newcommand{\pHe}{$\text{p}/\text{He}$\xspace}
\newcommand{\dpHe}{\text{p-He}\xspace}
\newcommand{\dpHeC}{p-He-C\xspace}

\begin{document} 

\begin{frontmatter}

  \title{Determination of the proton-to-helium ratio in cosmic rays at
    ultra-high energies from the tail of the \xmaxs distribution}
  
  \author[siegen,iteda]{A.~Yushkov\corref{ay}}
  \cortext[ay]{Corresponding author. Tel.: +49 271 740 3630}
  \ead{yushkov.alexey@gmail.com}
  \author[siegen]{M.~Risse}
  \author[siegen]{M.~Werner}
  \author[siegen]{J.~Krieg}

  \address[siegen]{Universit\"at Siegen, Department Physik, Walter-Flex-Str. 3, Siegen 57068}
  \address[iteda]{ITeDA (CNEA, CONICET, UNSAM),  Av. Gral. Paz 1499,
    Buenos Aires 1650}
\begin{abstract}
  We present a method to determine the proton-to-helium ratio in
  cosmic rays at ultra-high energies. It makes use of the exponential
  slope, $\Lambda$, of the tail of the \xmaxs distribution measured by
  an air shower experiment. The method is quite robust with respect to
  uncertainties from modeling hadronic interactions and to systematic
  errors on \xmaxs and energy, and to the possible presence of primary
  nuclei heavier than helium. Obtaining the proton-to-helium ratio
  with air shower experiments would be a remarkable achievement.

  To quantify the applicability of a particular mass-sensitive
  variable for mass composition analysis despite hadronic
  uncertainties we introduce as a metric the `analysis indicator' and
  find an improved performance of the $\Lambda$ method compared to
  other variables currently used in the literature.  The fraction of
  events in the tail of the \xmaxs distribution can provide additional
  information on the presence of nuclei heavier than helium in the
  primary beam.

\end{abstract}


\end{frontmatter}

\section{Introduction}
While the composition is known to be a key to understanding the origin
of ultra-high energy cosmic rays its determination by air shower
experiments is a challenge (for a review see
e.g.~\cite{kampert_mass2012}).  Estimations of the primary mass based
on observables such as the depth of shower maximum \xmaxs, particle
arrival times or muon content suffer from significant uncertainties in
the description of high-energy hadronic interactions. In this paper we
show that the exponential slope, $\Lambda$, of the tail of the \xmaxs
distribution can be used for the determination of the proton-to-helium
ratio, \pHe, in the primary beam and that $\Lambda$ and the inferred
\pHe are only weakly dependent on details of hadronic interactions and
experimental systematic uncertainties.  Thus, even with the indirect
observations performed by air shower experiments it becomes possible
to obtain a robust measure of an important detail of the cosmic ray
composition. Knowledge of \pHe can be used to constrain astrophysical
models of the origin of cosmic rays.

The exponential slope $\Lambda$ of the tail of the $X_\text{max}$
distribution was extensively studied and used with respect to the
determination of the proton-air interaction cross-section in an energy
region where the primary composition appears proton-dominated
(see~\cite{prl2012_cross,ulrich_icrc2015,TA_CX2015} and references
therein). This is due to the fact that the tail of the \xmaxs
distribution is intimately linked to the interaction cross-section of
the first interaction. One possible choice by such measurements was to
define the fit range so that it would contain 20\% of the deepest
events~\cite{prl2012_cross,ulrich_icrc2015}. The contamination of the
primary beam with photons and helium nuclei was the main source of the
systematic uncertainties on the proton-air cross-section. In the
present work this sensitivity of $\Lambda$ to the primary mass
composition is the very subject of the study. We define $\Lambda$ in a
different way, using a pre-defined fit range satisfying the condition
to be almost free from the contamination of nuclei heavier than
helium. This makes $\Lambda$ sensitive predominantly to \pHe in the
primary beam and provides a possibility for estimating \pHe in air
shower experiments.

Current best constraints on \pHe at ultra-high energies
$\gtrsim10^{18}$~eV can be deduced using the hadronic interaction
models EPOS-LHC~\cite{epos_muprod2008,epos_paris2009} and
QGSJet-II.04~\cite{qgsjetii} from mass composition fits of full \xmaxs
distributions measured at the Pierre Auger
Observatory~\cite{longXmax_fits2014}. Since the covariance matrices
for the fitted nuclei fractions are not available an estimation of the
upper limits on \pHe from this data with account of systematic and
statistical uncertainties can be done by taking the upper value of the
proton fraction $f_{\rm p}+\Delta f_{\rm p}({\rm stat+syst})$ and
dividing it by the lower value of the helium fraction $f_{\rm
  He}-\Delta f_{\rm He}({\rm stat+syst})$ (for the lower limits on
\pHe one has $(f_{\rm p}-\Delta f_{\rm p})/(f_{\rm He}+\Delta f_{\rm
  He})$). For the energy range \logenr{17.8}{18.4} where the fits
indicate the presence of a large fraction of protons and in case of
EPOS-LHC a very low fraction of helium, one gets
$\pHe(\text{QGSJet-II.04})=1.8\substack{+6.1 \\ -0.6}~({\rm
  stat+syst})$ and $\pHe(\text{EPOS-LHC})=22.1\substack{+142
  \\ -17.4}~({\rm stat+syst})$. Even when comparing just these two
post-LHC event generators, the central values of \pHe differ by
around a factor 10. Taking into account the uncertainties it is
possible to set only a lower limit $\pHe\gtrsim1$. For higher energies
\logenr{18.4}{19.4} the corresponding values
$\pHe(\text{QGSJet-II.04})=1.3\substack{+4.6 \\ -0.9}~({\rm
  stat+syst})$ and $\pHe(\text{EPOS-LHC})=1.7\substack{+5.9
  \\ -1.1}~({\rm stat+syst})$ cover a range from helium dominance to
helium being almost absent. These uncertainties may get even larger
when other hadronic models will be used, like the new version of
Sibyll~\cite{sibyll23_2015} producing showers with deeper \xmaxs
compared QGSJet-II.04 and EPOS-LHC. The $\Lambda$ method proposed in
this paper can considerably improve the ability to determine \pHe. The
slope $\Lambda$ is weakly sensitive to the variations of \xmaxs
distribution properties (mean $\langle\xmaxs\rangle$ and width
$\sigma(\xmaxs)$) which are influenced by shower development and
fluctuations and drive the fits to the full \xmaxs distributions, and
thus to (i) variations of hadronic interaction parameters other than
the cross-section and to (ii) the systematic uncertainties on \xmaxs
and energy, as will be discussed below. The use of $\Lambda$ is hence
expected to provide better and more robust constraints on \pHe
compared to fitting of the full \xmaxs distributions.

Due to our definition of the \xmaxs fit range, the fraction of events
in the fit range becomes a primary mass sensitive parameter itself. We
show that this fraction can be useful for additionally constraining
the abundances of nuclei heavier than helium. Having introduced more
variables in addition to a number of existing ones based on \xmaxs,
muon content and related observables (signal risetime,
radius of curvature etc.), one would like to compare their performance
for mass composition analysis in particular with respect to
uncertainties from modeling hadronic interactions.  We propose to use
the ratio $\Delta X({\rm p-He})/\Delta X(\text{h})$ of the difference
between proton and helium for some variable $X$ to the uncertainties
$\Delta X(\text{h})$ on this variable due to the limited knowledge of
hadronic interaction properties. We call this ratio an `analysis
indicator'. The analysis indicator for the exponential slope $\Lambda$
turns out to be significantly larger than for other mass-sensitive
variables such as $\langle\xmaxs\rangle$ or average number of muons.

All simulations are performed with CONEX~\cite{conex_2006,conex_pylos}
for the hadronic interaction models EPOS-LHC and QGSJet-II.04. For
each primary particle type (proton, helium, carbon, iron) and energy
($\loge=18.0,18.5$ and $19.0$) around $10^{6}$ showers are
produced. To imitate the detector resolution effect, each \xmaxs value
is smeared by adding a Gaussian distributed random variable with
$\sigma=20$~\gsm, a typical value for current shower experiments using
the fluorescence technique~\cite{WGmass_UHECR2014}.

In addition to showers with fixed energies, we simulated $10^5$
EPOS-LHC proton showers from an energy spectrum with $\gamma=3.23$ in
the energy range \logenr{17.8}{18.2} to confirm that a typical energy
distribution does not affect the determination of the value of
$\Lambda$ due to its weak energy dependence and that the simulations
for fixed energies provide the necessary precision.

\section{Method for the determination of 
  \pHe in the primary beam}

A main feature of our approach is the definition of the fit range,
given a measured \xmaxs distribution.  In the cross-section analysis,
where the fit range is determined by requiring it to contain the
deepest $\eta=20\%$ of events, for a simplified example of a pure carbon
composition one will get $\Lambda_\eta$ corresponding to the
carbon-air cross-section. For mixed compositions $\Lambda_\eta$ takes
values from the spectrum of possibilities from proton to heavier
nuclei and intermediate values are degenerate, i.e. the same value of
$\Lambda_\eta$ corresponds to a number of different mixed
compositions.

Instead, we define a lower limit of the \xmaxs fit range using the
requirement that only $\approx0.5\%$ of the carbon-initiated showers
survive. This way, a very small contamination from nuclei heavier than
helium can be achieved. In this paper we use the lower limit value
originating from QGSJet-II.04 which is $5-7$~\gsm\ larger compared to
the corresponding value for EPOS-LHC (thus the fraction of the
surviving carbon showers for EPOS-LHC is slightly below 0.5\%). The
lower limit of the fit range defined this way is $809+48(\log(E/{\rm
  eV})-18)$~\gsm. The values of $\Lambda$ change within
$\approx1-2$~\gsm\ when the width of the fit range is varied within
$60-150$~\gsm, further we use a width of $100$~\gsm. For application
to data the width could possibly be further optimized depending on the
characteristics of the specific experiment and the available number of
events.

Due to our definition, in case of a composition devoid of protons and
helium (such as the example of a pure carbon composition) there will
be too few events in the fit range to get a statistically reliable
estimation of $\Lambda$ (the usage of the fraction of events in the
fit range for composition studies is inspected in Sec.~3). For mixed
compositions, $\Lambda$ in our method is sensitive mainly to the
relative abundances of protons and helium in the primary beam, and
would take values between the $\Lambda$ values for pure proton and
pure helium as will be shown below.

In Fig.~\ref{fig:xmaxdist} examples of proton, helium and carbon
\xmaxs distributions are shown (left panel). There is a clear
difference in the slopes between protons and helium. In the right
panel of Fig.~\ref{fig:xmaxdist} fits of the function
$\exp(-\xmaxs/\Lambda)$ to the tail of proton EPOS-LHC \xmaxs
distribution are given for energies of \logen{18.0} and 19.0.  We use
an unbinned maximum likelihood method, the binned histograms in the
plots are given for visualization purpose.

\begin{figure}[!h]
\centering\includegraphics[width=0.65\textwidth]{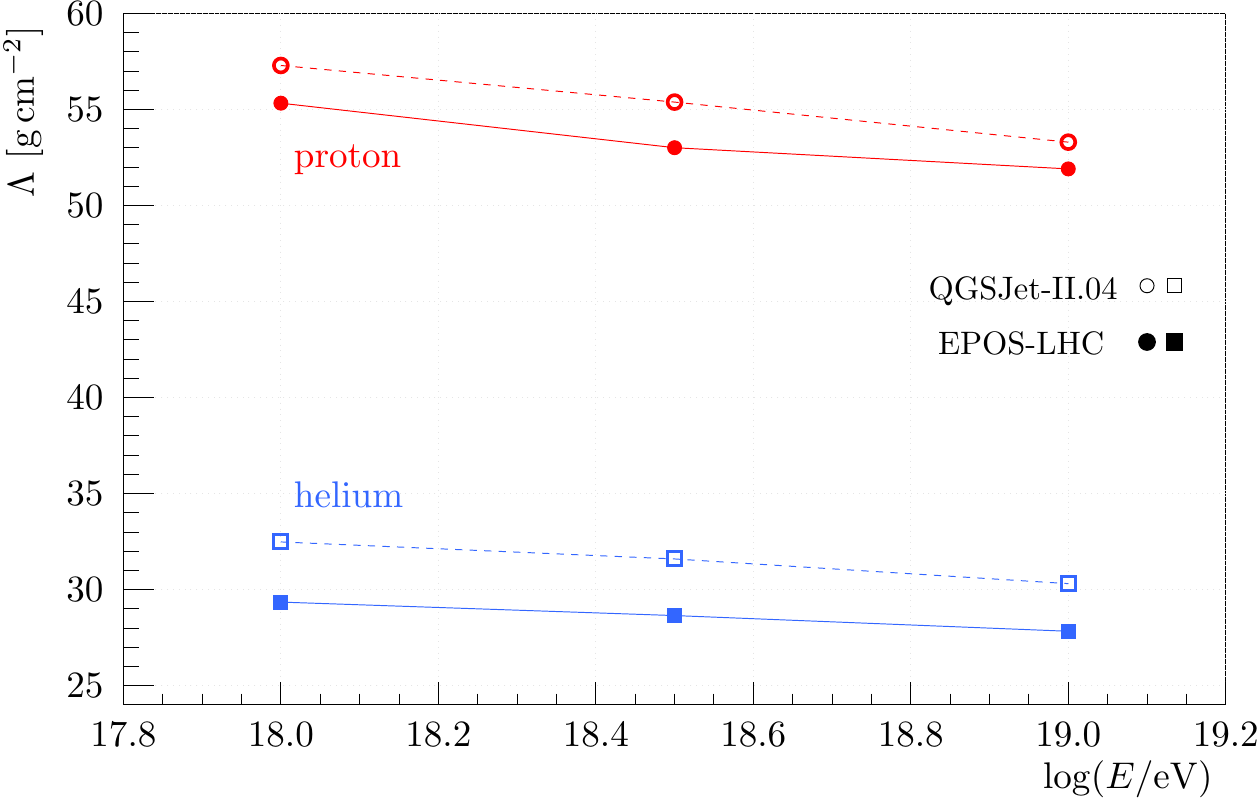} 
\caption{Energy dependence of the exponential slopes $\Lambda$ of the
  tails of proton and helium \xmaxs distributions for two hadronic
  interaction models.}
\label{fig:lambda}
\end{figure}

The results for $\Lambda$ for all 3 primary energies and both
interaction models are summarized in Fig.~\ref{fig:lambda}.  The value
of $\Lambda$ for proton is $\approx20-25$~\gsm larger than that for
helium (due to the significantly larger helium-air cross-section, see
e.g.~\cite{wibig_cx1998}). This difference is much larger than the
difference of $\approx2-3$~\gsm due to the hadronic model
uncertainties, indicating an excellent suitability of the
$\Lambda$ method for mass composition analysis with only a minor
dependence on details of the hadronic interaction models.

\begin{figure}[!h]
\includegraphics[width=0.5\textwidth]{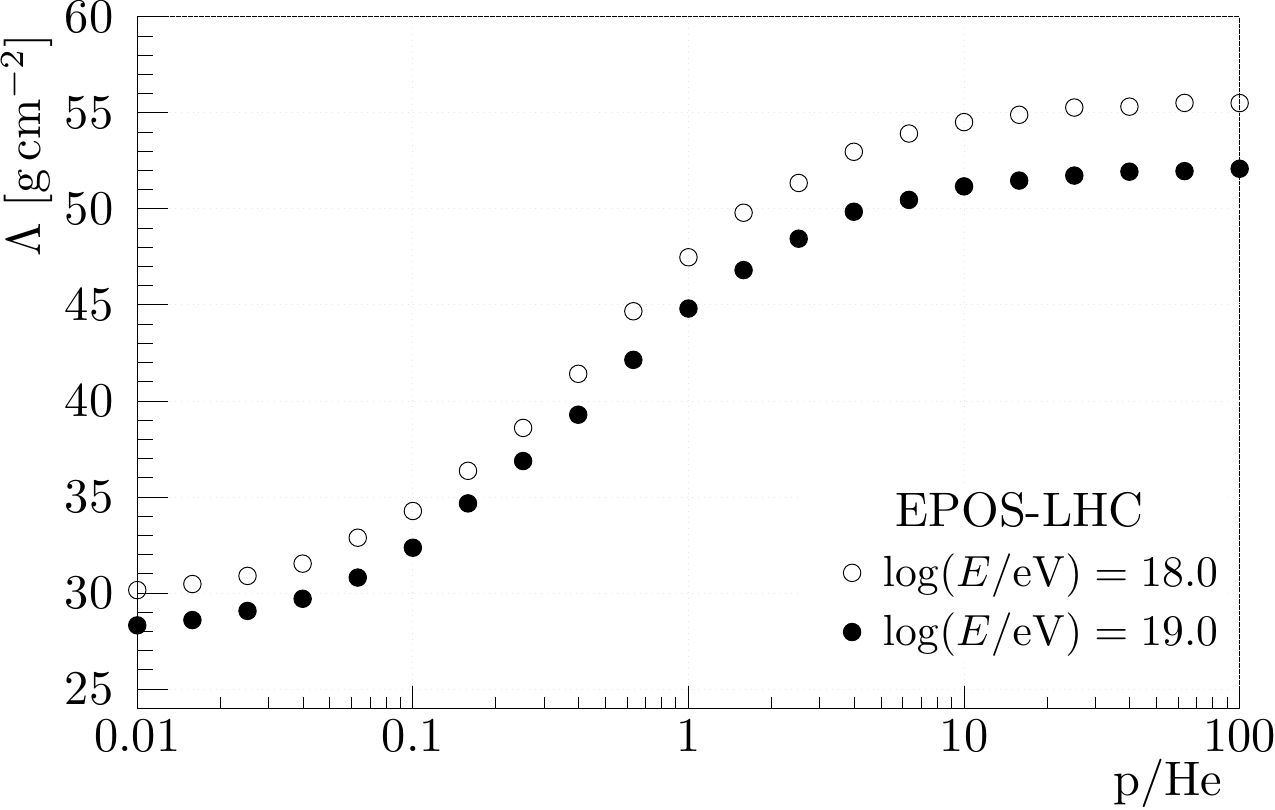} 
\includegraphics[width=0.5\textwidth]{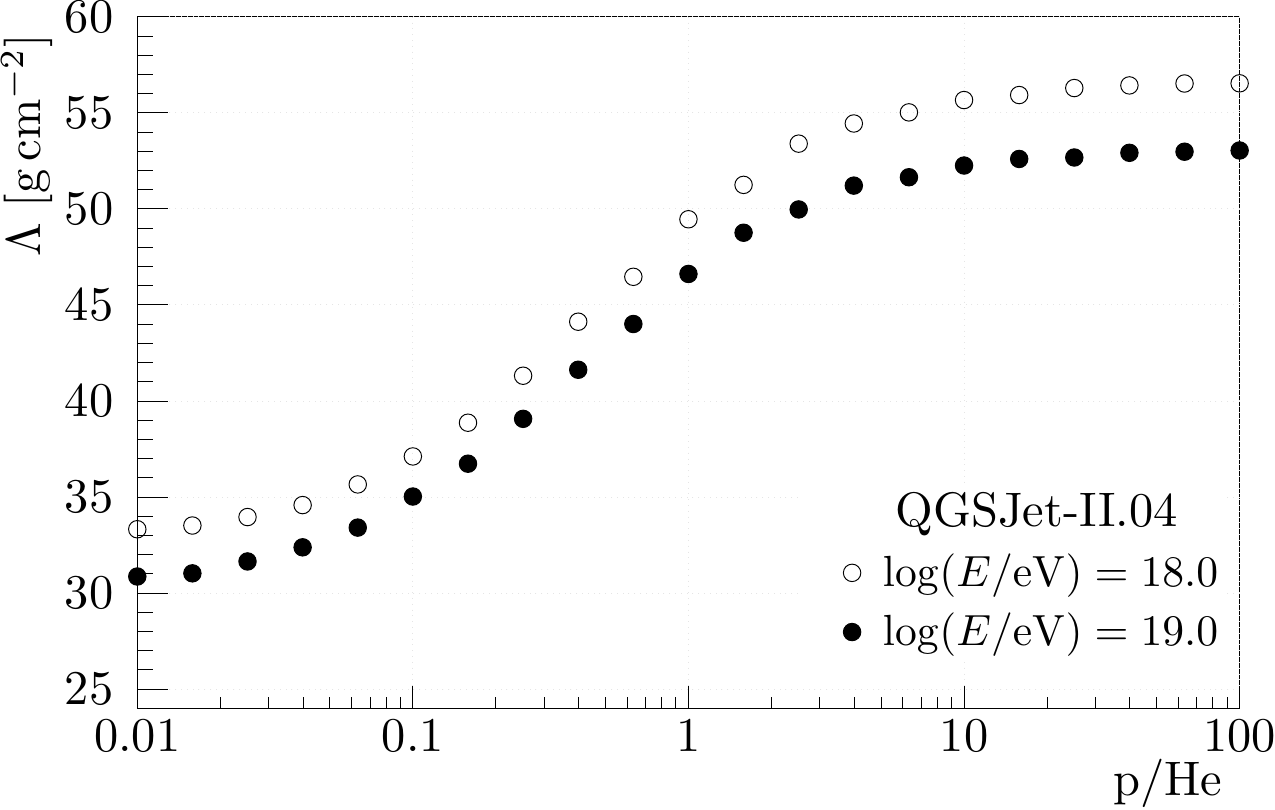}\\
\caption{Dependence of $\Lambda$ on \pHe for EPOS-LHC (left) and
  QGSJet-II.04 (right) in two-component \dpHe mixtures at
  \logen{18.0} and \logen{19.0}.}
\label{fig:pHe}
\end{figure}

In Fig.~\ref{fig:pHe} we show the dependence of $\Lambda$ on \pHe for
two-component \dpHe mixtures. An accurate determination of \pHe is
possible in a range $0.1\lesssim {\rm p/He} \lesssim 3$ where
$\Lambda$ and \pHe are strongly correlated. Outside this range
limits ${\rm p/He} \lesssim 0.1$ or ${\rm p/He} \gtrsim 3$ can be set.

\begin{figure}[!h]
\includegraphics[width=0.5\textwidth]{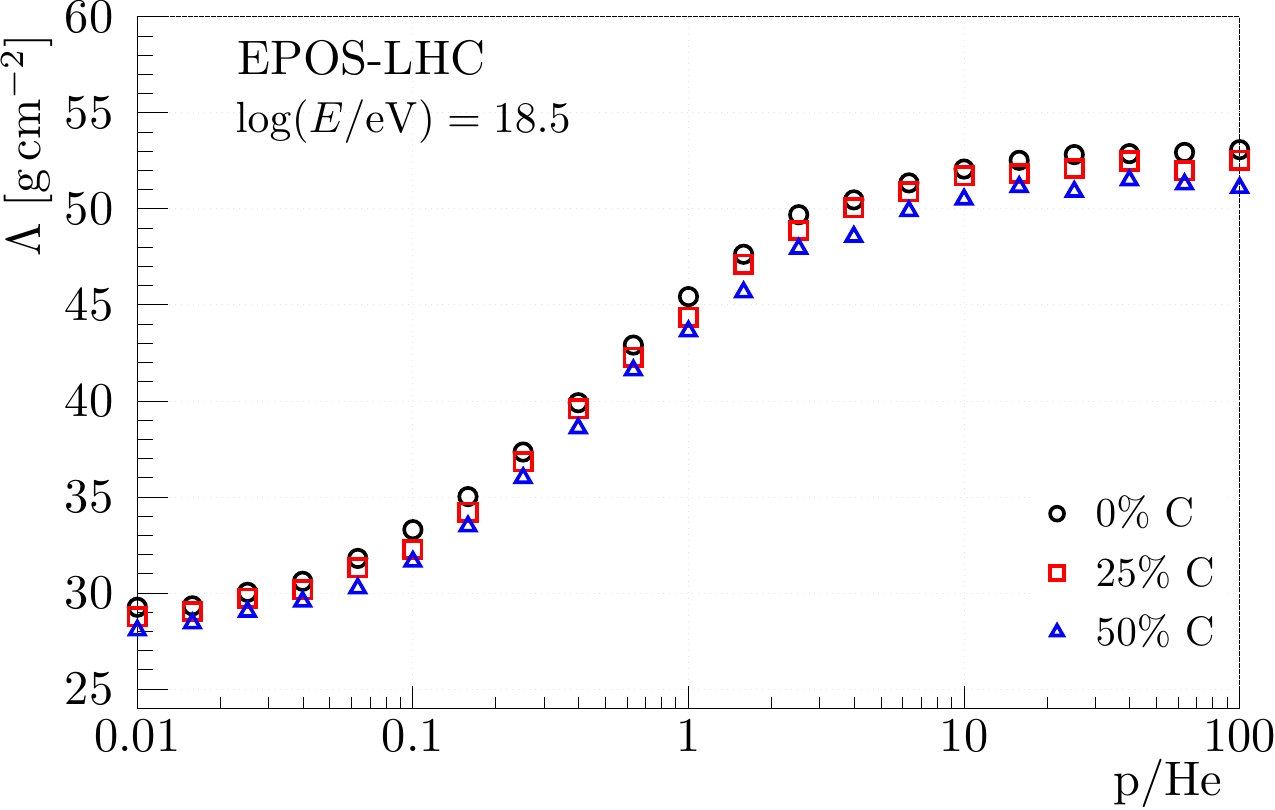} 
\includegraphics[width=0.5\textwidth]{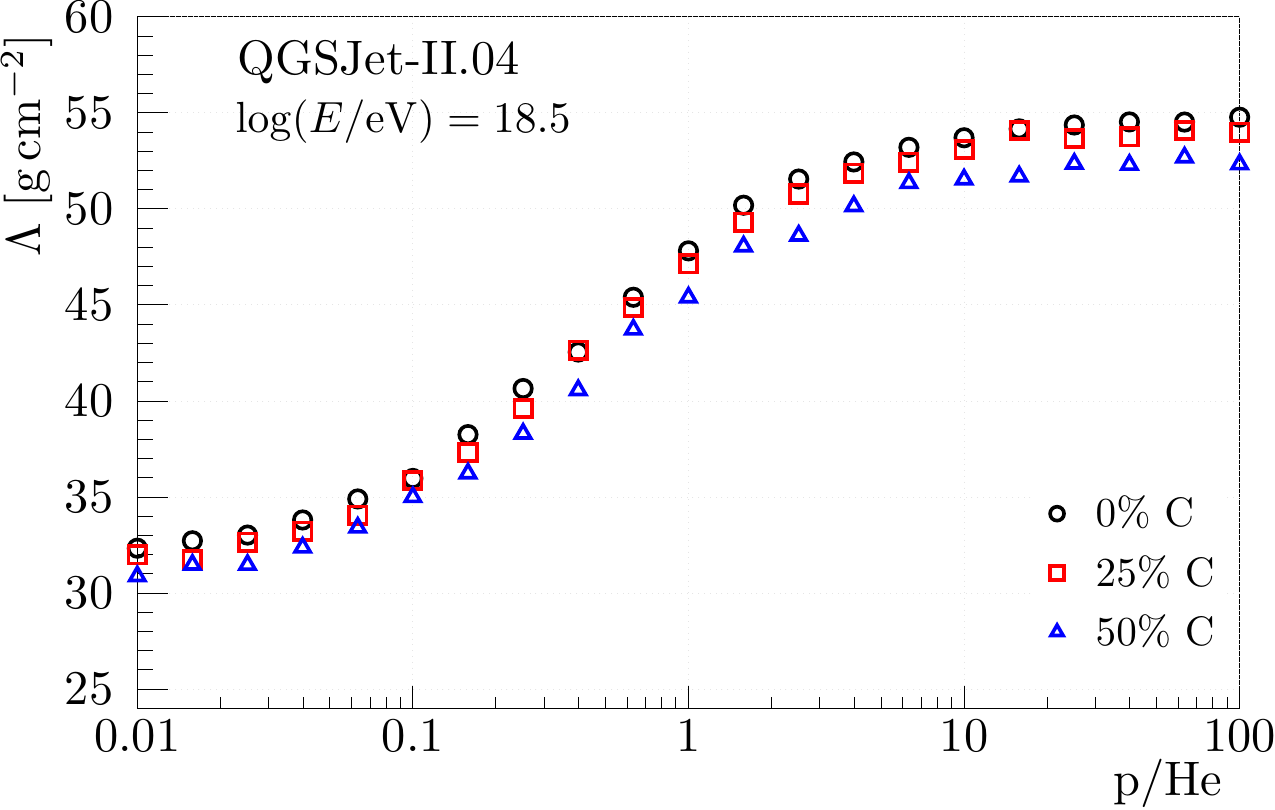}\\
\caption{Dependence of $\Lambda$ on \pHe for EPOS-LHC (left) and
  QGSJet-II.04 (right) in two-component \dpHe mixtures ($0\%$~C)
  compared to the dependence in three-component \dpHeC mixtures with
  carbon fractions of 25\% and 50\% for \logen{18.5}.}
\label{fig:pHeC}
\end{figure}

Next, we investigate the impact of primary nuclei heavier than
helium. We focus here on adding carbon ($A = 12$) and note that the
effect of even heavier nuclei such as oxygen or iron is smaller than
that of carbon. In Fig.~\ref{fig:pHeC} the values of $\Lambda$ for
\dpHe mixtures are compared to the values for three-component mixtures
\dpHeC with carbon fractions of 25\% and 50\% in the primary beam. As
expected from the definition of the \xmaxs fit range, the results on
$\Lambda$ (and thus on primary \pHe) are very robust. Specifically, a
contamination with carbon of even 50\% leads to a shift of $\Lambda$
of $2-3$~\gsm only.  A determination of $\Lambda$ with an uncertainty
of few \gsm\ (e.g. for \dpHe samples of 5000 events in total, the
relative statistical uncertainty is
$\Delta\Lambda/\Lambda\approx(8-10)\%$,
c.f. also~\cite{ulrich_icrc2015}) in several energy bins allows one to
reconstruct the evolution of \pHe without a significant bias coming
from variations in the fractions of heavier elements.

The value of $\Lambda$ is quite robust with respect to variations of
the mean $\langle\xmaxs\rangle$ and the width $\sigma(\xmaxs)$ of
\xmaxs distributions, and thus to experimental systematic
uncertainties on \xmaxs or to the variations of hadronic interaction
parameters other than the cross-section. For example, the smearing
with a Gaussian distributed random variable with $\sigma=20$~\gsm that
we apply to the CONEX \xmaxs values, changes $\Lambda$ within
1.5~\gsm. The systematic uncertainty on \xmaxs reported by the Auger
Observatory~\cite{longXmax2014} is smaller than 10~\gsm. Shifts of the
simulated \xmaxs values by $\pm10$~\gsm produce changes in the values
of $\Lambda$ within $1-2$~\gsm. The systematic uncertainty on energy
(measured with the fluorescence technique) of 14\% corresponds to
$\approx3.5$~\gsm uncertainty on \xmaxs and thus practically does not
affect the values of $\Lambda$.

For application to real data, it is worth noting that special care has
to be taken to achieve an unbiased data set and, thus, an unbiased
extraction of $\Lambda$. This may involve a dedicated event selection
based on shower geometries such as performed, for instance,
in~\cite{prl2012_cross,ulrich_icrc2015}.

\begin{figure}[!h]
\includegraphics[width=0.5\textwidth]{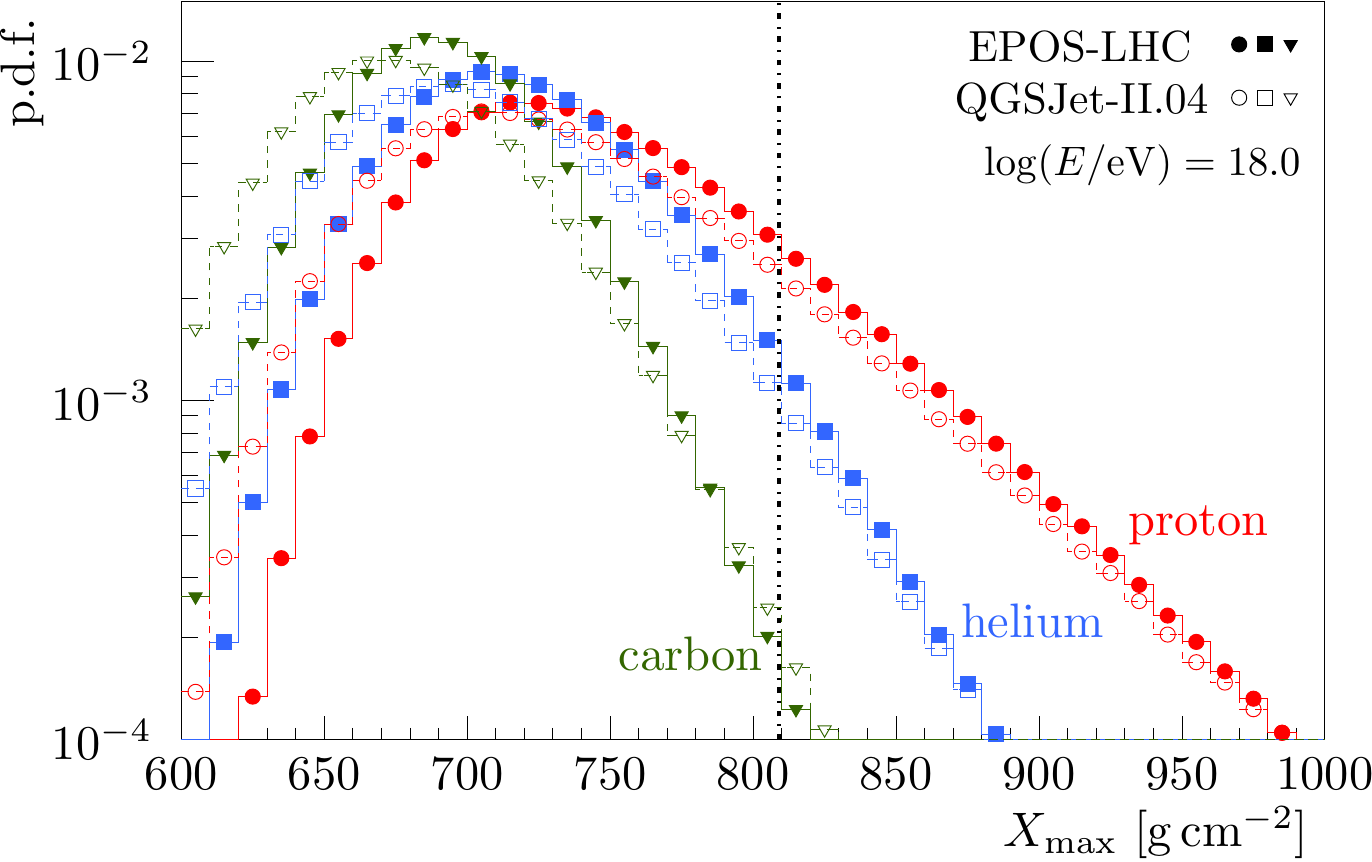} 
\includegraphics[width=0.5\textwidth]{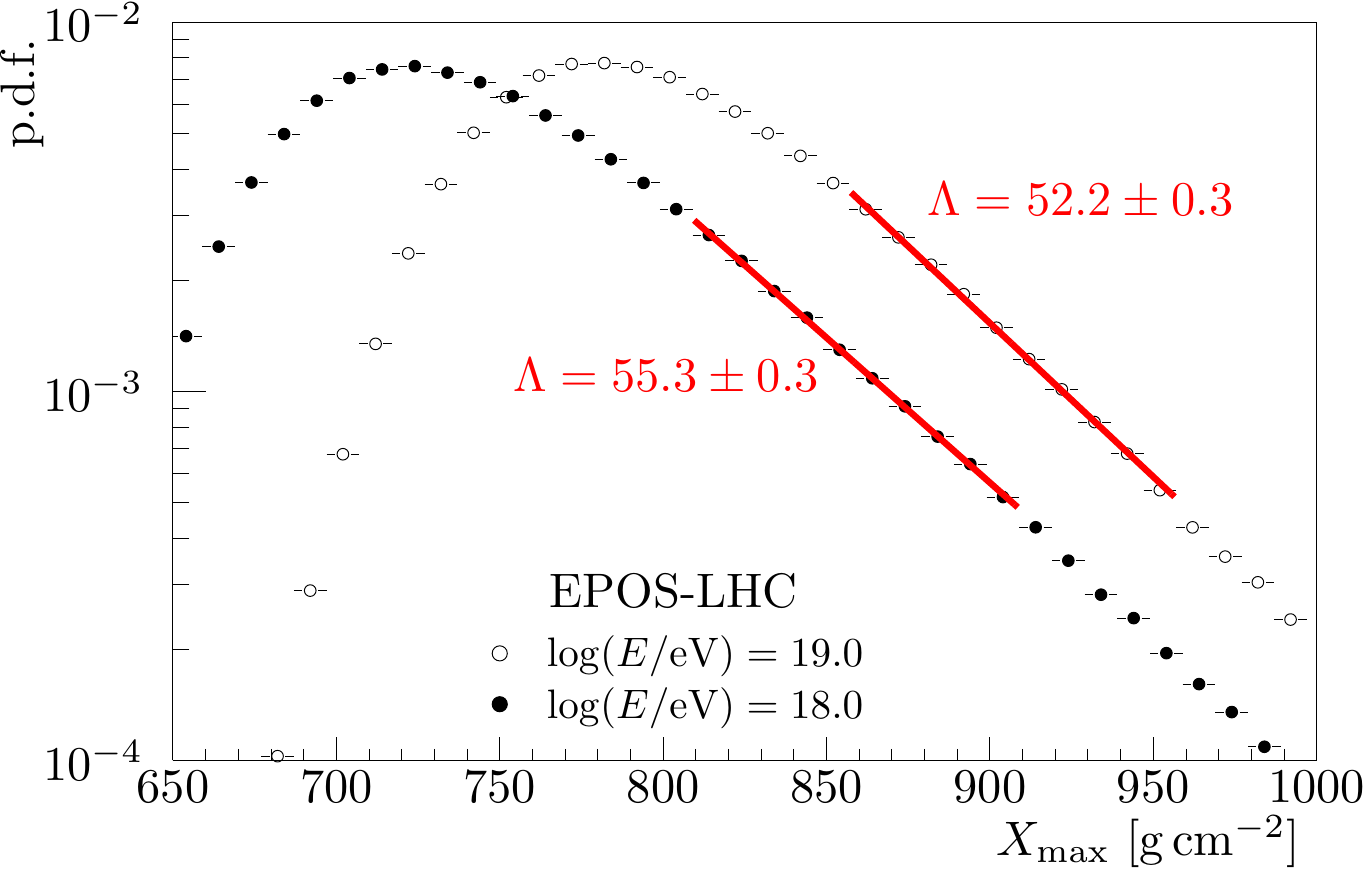} 
\caption{Left: \xmaxs distributions for proton, helium and carbon for
  EPOS-LHC and QGSJet-II.04 at \logen{18.0}. The vertical line
  indicates the lower limit of the fit range. Right: examples of
  $\exp(-\xmaxs/\Lambda)$ fits of the tails of the \xmaxs
  distributions for EPOS-LHC proton at \logen{18.0} and
  19.0. $\Lambda$ values are in units of $\mathrm{g\,cm^{-2}}$.}
\label{fig:xmaxdist}
\end{figure}

\begin{figure}
\centering\includegraphics[width=0.65\textwidth]{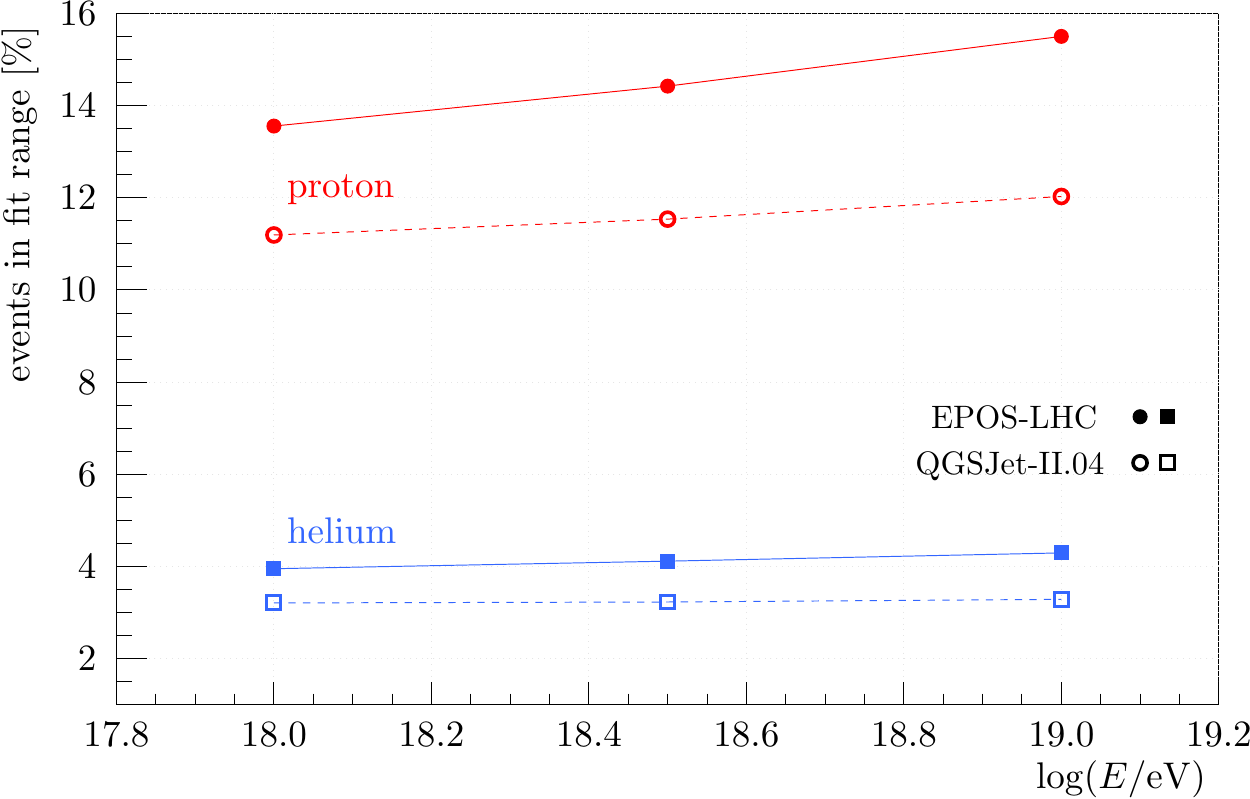} 
\caption{Fractions of events in the fit range for proton and helium.}
\label{fig:fractions}
\end{figure}

\begin{figure}
\includegraphics[width=0.5\textwidth]{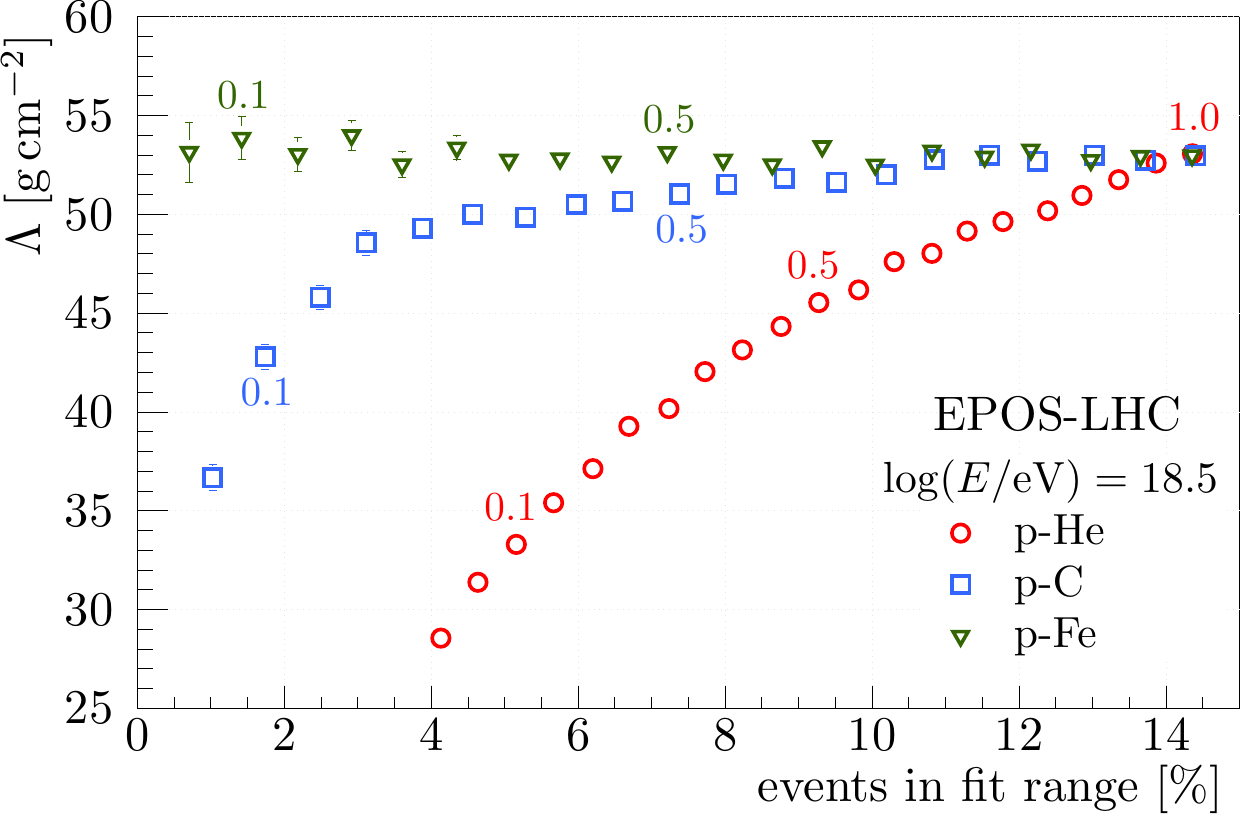} 
\includegraphics[width=0.5\textwidth]{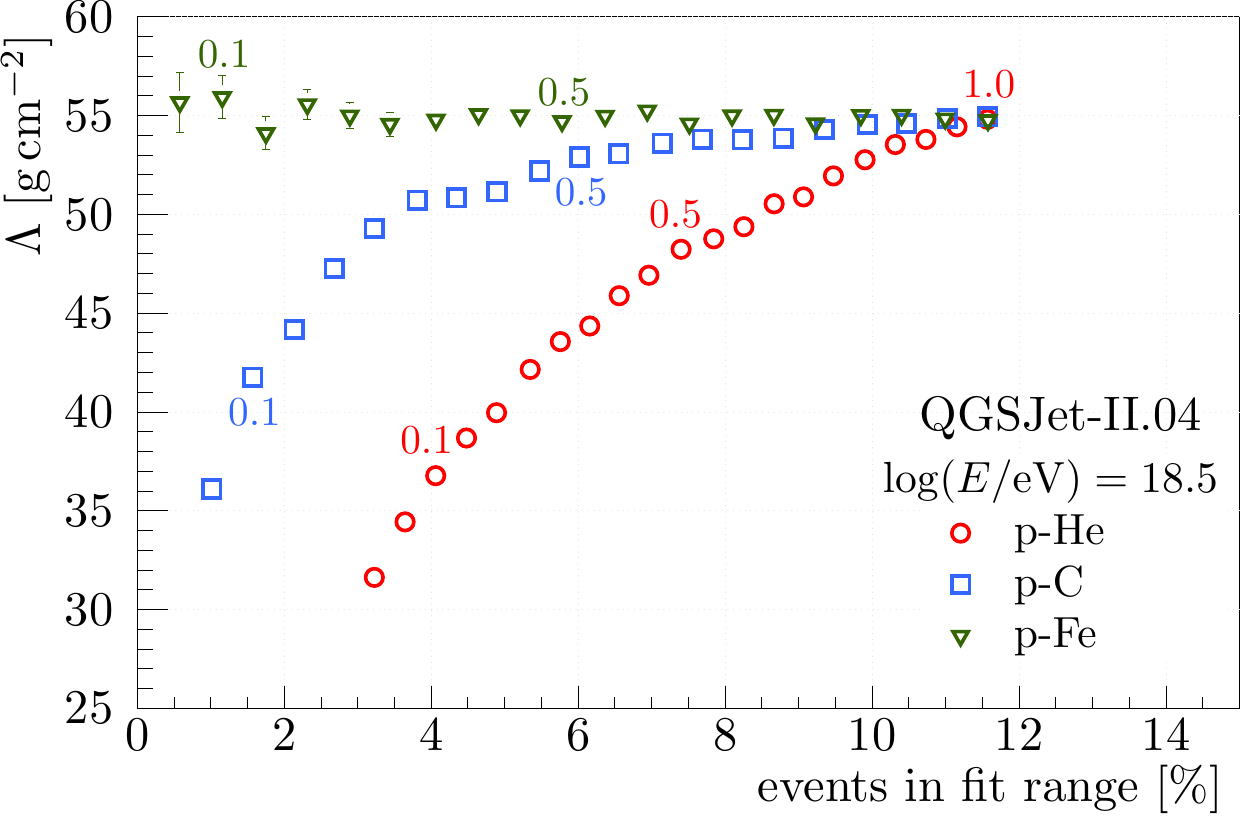} 
\caption{Exponential slope vs. fraction of events in the fit range for
  EPOS-LHC (left) and QGSJet-II.04 (right) for p-He, p-C and p-Fe
  mixtures at \logen{18.5}. The relative fraction of primary particles
  changes with a step of $0.05$. Numbers $0.1, 0.5, 1.0$ near the
  graphs indicate the proton fraction in the samples. For pure carbon
  and iron samples there are too few events in the fit range to obtain
  a reliable estimation of $\Lambda$, thus there are no corresponding
  points in the plots.}
\label{fig:lambda_frac}
\end{figure}

\section{Fraction of events in the fit range}

For the proposed $\Lambda$ definition the fraction of events in the
fit range becomes itself a mass-sensitive variable. For both models
the fractions of the carbon \xmaxs distribution in the fit range are
below $0.5\%$. The fractions of the distributions for protons and
helium nuclei are shown in Fig.~\ref{fig:fractions}, ranging from
$11.5\%-15.5\%$ for protons and $3.2\%-4.3\%$ for helium depending on
energy and interaction model. Thus, one can see a good separation
between protons and helium also in this variable though differences
between the hadronic interaction models are larger compared to the
case of $\Lambda$ (cf. Fig.~\ref{fig:lambda}).  While with $\Lambda$
one can extract the primary \pHe, the fraction of events can give
additional information about the presence of heavier nuclei in the
primary beam. Therefore one can combine the information from both
variables in a single plot, as shown in
Fig.~\ref{fig:lambda_frac}. One can see that for proton-carbon
mixtures $\Lambda$ decreases only slightly ($<2$~\gsm) for $50\%$
fractions of carbon (c.f. Fig.~\ref{fig:pHeC}). In case the observed
$\Lambda$ is compatible with the expectation for protons, the fraction
of events can be used for constraining the absolute proton
abundance. For example, a fraction of events $<6\%$ corresponds to a
proton fraction $\lesssim50\%$ for both interaction models and
all compositions. In case both $\Lambda$ and the fraction of events
are compatible with the values for proton-helium mixtures, one can
conclude that the admixture of heavier elements is small. Dilution of
the primary beam with heavier elements would be seen as a horizontal
shift of the data point to the left from the proton-helium graph,
towards smaller values of the fraction of events in the fit range. Any
data points in the area above the proton-only value of $\Lambda$ or to
the right of the proton-helium line would serve as an indication of
shortcomings in the particular interaction model.

Compared to the fraction of events in the fit range, $\Lambda$ is a
more robust variable with respect to the uncertainties in the
interaction models, being sensitive essentially to the cross-section
only. The fraction of events in the fit range is influenced both by
the position of the shower maximum and by the width of the $\xmaxs$
distributions. Specifically, shifting \xmaxs values in accordance with
systematic uncertainties of the Auger Observatory by $\pm10$~\gsm one
gets the following variations in the fraction of events in the fit
range at \logen{18.5}: $14.4\substack{+2.9\\ -2.4}\%$ (protons) and
$4.1\substack{+1.6\\ -1.2}\%$ (helium) for EPOS-LHC; and
$11.5\substack{+2.3\\ -1.9}\%$ (protons) and
$3.2\substack{+1.2\\ -0.8}\%$ (helium) for QGSJet-II.04. A metric to
quantify the impact of the hadronic interaction uncertainties for
variables such as $\Lambda$ and fraction of events is introduced in
the next section.

\section{Analysis indicator}

Denoting for some variable $X$ the uncertainty due to the hadronic
(h) interaction models as $\Delta X(\text{h})$ and the difference between
values of $X$ for protons and helium as $\Delta X({\rm p-He})$ one can
define the following quantity:

$$
\iah{X}=\frac{\Delta X({\rm p-He})}{\Delta X(\text{h})}.
$$

We propose to name it `analysis indicator': values $\iah{X}\gg1$ (that
is, $\Delta X({\rm p-He})\gg\Delta X(\text{h})$) indicate that $X$ is
excellent for mass composition (label `$A$' in \iah{X}) analysis,
while for $\iah{X}\ll1$ the variable is good for constraining
interaction models (label `h'). Thus $I^A_\text{h}$ provides a metric
to judge on the reliability of a variable to determine the composition
or hadronic parameters. For many variables it may be close to 1,
indicating that they are of limited use as an isolated parameter for
distinguishing changes in the mass composition from changes or
uncertainties in hadronic interactions. Let us consider several
examples.

The depth of the shower maximum is known as a good shower observable
for estimating the mass composition. A main variable extracted from a
set of measured \xmaxs values is the mean $\langle\xmaxs\rangle$. At
energies around 10~EeV the difference between protons and helium
in $\langle \xmaxs\rangle$ is around 35~\gsm\ and the uncertainty due to
interaction models is around 20~\gsm~(see e.g.~\cite{longXmax2014})
giving a quite large value of the analysis indicator $\iah{\langle
  \xmaxs\rangle}\approx35~[\gsm]/20~[\gsm]\approx1.7$.

An opposite example can be given by variables related to the muon
shower content. For the muon production depth~\cite{mpd2014} the
analysis indicator is\footnote{$\Delta \langle
  X_\text{max}^\mu\rangle(p-\text{He})$ is obtained using $\Delta
  \langle X_\text{max}^\mu\rangle(p-\text{Fe})\times\ln4/\ln56$}
$\iah{\langle X_\text{max}^\mu\rangle}\approx25~[\gsm]/50~[\gsm]=0.5.$
Similarly for the number of muons in very inclined air
showers\footnote{$R_\mu$ is the ratio of the total number of muons in
  simulations or data to a number of muons in a typical proton shower
  at 10~EeV produced with QGSJet-II.03} $R_\mu$~\cite{has_muons2015}
one gets $\iah{\langle\ln R_\mu\rangle}\approx0.1/0.2=0.5.$
Uncertainties on these variables from the description of hadronic
interactions are limiting their usage for composition
analysis. Results on \meanlnA from $R_\mu$ and $X_\text{max}^\mu$
leading to masses heavier than iron~\cite{mpd2014,has_muons2015} may
be used to restrict interaction models.

For $\Lambda$ the uncertainties due to interaction models are small
$\approx2-3$~\gsm\ and the difference between protons and helium is
around $20$~\gsm (Fig.~\ref{fig:lambda}), thus
$\iah{\Lambda}\approx7-10$. Allowing for an uncertainty (statistical and
systematic) of approximately 10\% on the proton-air
cross-section~\cite{prl2012_cross} which corresponds to an uncertainty
of $\approx5-6~\gsm$ on $\Lambda$, the analysis indicator is still
large $\iah{\Lambda}\approx20~[\gsm]/6~[\gsm]\approx3.3$. This
indicates that the $\Lambda$ method is suitable for differentiating
between primaries even as close as proton and helium.

For the fraction of events in the fit range the analysis indicator is
$\iah{{\rm fraction}}\approx7\%/3\%\approx2.3$
(Fig.~\ref{fig:fractions}) which is clearly smaller compared to
$\iah{\Lambda}$ but still indicates a good applicability for
composition studies.

\section{Discussion}

We introduced a method to determine \pHe in primary cosmic rays at
ultra-high-energies from the slope of the exponential fit of the tail
of the \xmaxs distribution. The method is robust with regard to a
possible contamination of the primary beam with nuclei heavier than
helium, to experimental systematic uncertainties on \xmaxs and energy,
and -- importantly -- to uncertainties in hadronic models being
sensitive mainly to the interaction cross-section. A solid
determination of \pHe with air shower experiments will be a remarkable
achievement. The fraction of events in the fit range is proposed as an
additional mass sensitive variable that can be used to constrain the
abundance of nuclei heavier than helium in the primary beam.

As explained in the introduction, for the determination of \pHe the
use of the full $X_{\rm max}$ distribution to fit the cosmic ray
composition is not superior or equivalent to the $\Lambda$ method
presented here which exploits only the tail of the distribution. This
can be understood because the absolute \xmaxs values (and related
$\langle\xmaxs\rangle$) do not have such a direct relation to the
primary \pHe as has $\Lambda$.  Moreover, the interpretation of the
absolute \xmaxs values is much stronger affected by experimental
systematics e.g. in \xmaxs, by the (uncertain) presence of heavier
nuclei, and, in particular, by hadronic model uncertainties
($\iah{\Lambda}>\iah{\langle\xmaxs\rangle}$). Since, however, the absolute \xmaxs
values are a main driver when fitting the full \xmaxs distribution,
the primary \pHe from elemental fractions derived this way suffers
from larger uncertainty compared to exploiting the direct correlation
between $\Lambda$ and primary \pHe. While the $\Lambda$ method
provides only partial but not full information on composition such as
abundances of all elements or mass groups (as fitting the full \xmaxs
distribution does), it focuses on that part of the data which allows
us to draw robust conclusions about the mass composition, even if the
price to pay is to determine only one of its specific aspects, namely
in this case \pHe.

In the paper on proton-air cross-section measurement by the Pierre
Auger Observatory~\cite{prl2012_cross} the fit of $\eta=20\%$ of the
deepest events with energies \logenr{18.0}{18.5} gave
$\Lambda_\eta=55.8\pm2.3\,({\rm stat})\pm1.6\,({\rm syst})$~\gsm. A
naive, straightforward comparison of this value with the values found
in the current paper (Fig.~\ref{fig:pHe}) indicates that protons are
indeed existing and are more abundant than helium nuclei at these
energies, in agreement with the estimations
of~\cite{prl2012_cross}. However, such a comparison should be taken
with great care. The fit range in the cross-section analysis (both the
starting depth and the width) differs from the definition in the
current work, optimized for \pHe determination. Such a difference will
result in a different data sample for the \pHe analysis compared to
the data sample used for the cross-section analysis. As a result the
value of $\Lambda_\eta$ might differ as well both due to an earlier
start of the fit (closer to the maximum of the \xmaxs distribution)
and due to a possible larger contamination with nuclei heavier than
helium. Thus for a quantitative interpretation of data a proper
accounting for detector acceptance, event selection criteria and
definition of the fit range is needed.

In the method proposed in the current paper, an important
characteristic we exploit is the quasi-discreteness of light nuclei
($A<12$) in the cosmic ray beam with protons and helium being the
dominating light primary particles.  Other light nuclei are not
expected to significantly impact the relation between $\Lambda$ and
\pHe.  The primary abundance of the other light nuclei is expected to
be very small due to their relative instability compared to protons
and helium.  As confirmed by simulations of cosmic ray propagation
with CRPropa~\cite{crpropa2}, their primary abundance is suppressed by
factors $>10$ compared to the abundance of ($\rm
p+He$)~\cite{crpropa2,puget1976}.  Furthermore, elements heavier than
helium are additionally suppressed by our definition of the \xmaxs fit
range.  For instance, suppression factors due to the
fit range relative to protons (to helium) are about 8 (2.5) for
${}^7$Li and 20 (6) for ${}^{10}$B.

Finally, we also introduced a new quantity called `analysis indicator'
for the characterization of the performance of the mass-sensitive
variables for mass composition analysis. The value of the analysis
indicator for $\Lambda$ is at least twice as large as that for
$\langle\xmaxs\rangle$. Compared to $\langle\xmaxs\rangle$ or to an
analysis making use of absolute \xmaxs values, $\Lambda$ is more
robust with regard to experimental systematic uncertainties and to
uncertainties in the simulation of hadronic interactions since it
depends essentially only on the cross-section. Though the fraction of
events in the fit range is more sensitive to hadronic interaction
parameters compared to $\Lambda$, it still has a value of the analysis
indicator comparable to that for $\langle\xmaxs\rangle$. The analysis
indicator can provide a common base for comparing performances of the
different mass-sensitive variables for mass composition analysis or
for constraining hadronic interactions.

\section*{Acknowledgments}
The work is supported by the German Federal Ministry of Education and
Research (BMBF) and by the Helmholtz Alliance for Astroparticle
Physics (HAP).

We are very grateful to our colleagues and friends from the Pierre
Auger Collaboration for the invaluable experience that we acquired
from them during many years of the joint work.

We are grateful to Tobias Winchen for providing us with CRPropa
simulations. We thank the anonymous referees for their constructive comments.

\iftrue

\fi


\begin{thebibliography}{18}
\expandafter\ifx\csname natexlab\endcsname\relax\def\natexlab#1{#1}\fi
\providecommand{\url}[1]{\texttt{#1}}
\providecommand{\href}[2]{#2}
\providecommand{\path}[1]{#1}
\providecommand{\DOIprefix}{doi:}
\providecommand{\ArXivprefix}{arXiv:}
\providecommand{\URLprefix}{URL: }
\providecommand{\Pubmedprefix}{pmid:}
\providecommand{\doi}[1]{\href{http://dx.doi.org/#1}{\path{#1}}}
\providecommand{\Pubmed}[1]{\href{pmid:#1}{\path{#1}}}
\providecommand{\bibinfo}[2]{#2}
\ifx\xfnm\relax \def\xfnm[#1]{\unskip,\space#1}\fi
\bibitem[{Kampert and Unger(2012)}]{kampert_mass2012}
\bibinfo{author}{K.-H. Kampert}, \bibinfo{author}{M.~Unger},
\newblock \bibinfo{title}{{Measurements of the Cosmic Ray Composition with Air
  Shower Experiments}},
\newblock \bibinfo{journal}{Astropart.Phys.} \bibinfo{volume}{35}
  (\bibinfo{year}{2012}) \bibinfo{pages}{660--678}.
\bibitem[{Abreu et~al.(2012)}]{prl2012_cross}
\bibinfo{author}{P.~Abreu}, et~al.,
\newblock \bibinfo{title}{{Measurement of the proton-air cross-section at
  $\sqrt{s}=57$ TeV with the Pierre Auger Observatory}},
\newblock \bibinfo{journal}{Phys.Rev.Lett.} \bibinfo{volume}{109}
  (\bibinfo{year}{2012})
  \bibinfo{pages}{062002}. (\bibinfo{collaboration}{Pierre Auger Collaboration}).
\bibitem[{Ulrich et~al.(2015)}]{ulrich_icrc2015}
  \bibinfo{author}{R.~Ulrich}, et~al., \bibinfo{title}{{Extension of
      the measurement of the proton-air cross section with the Pierre
      Auger Observatory}}, in: Proceedings of 34th International
  Cosmic Ray Conference, the Hague,
  2015. \bibinfo{note}{arXiv:1509.03732}.
  (\bibinfo{collaboration}{Pierre Auger Collaboration}).
\bibitem[{Abbasi et~al.(2015)}]{TA_CX2015}
  \bibinfo{author}{R. Abbasi}, et~al., \newblock
  \bibinfo{title}{{Measurement of the proton-air cross section with
      Telescope Array’s Middle Drum detector and surface array in
      hybrid mode}}, \newblock \bibinfo{journal}{Phys. Rev.}
  \bibinfo{volume}{D92} (\bibinfo{year}{2015})
  \bibinfo{pages}{032007}. (\bibinfo{collaboration}{Telescope Array
    Collaboration}).
\bibitem[{Pierog and Werner(2008)}]{epos_muprod2008}
\bibinfo{author}{T.~Pierog}, \bibinfo{author}{K.~Werner},
\newblock \bibinfo{title}{{Muon Production in Extended Air Shower
  Simulations}},
\newblock \bibinfo{journal}{Phys. Rev. Lett.} \bibinfo{volume}{101}
  (\bibinfo{year}{2008}) \bibinfo{pages}{171101}.
\bibitem[{Pierog and Werner(2009)}]{epos_paris2009}
\bibinfo{author}{T.~Pierog}, \bibinfo{author}{K.~Werner},
\newblock \bibinfo{title}{{EPOS Model and Ultra High Energy Cosmic Rays}},
\newblock \bibinfo{journal}{Nucl. Phys. Proc. Suppl.} \bibinfo{volume}{196}
  (\bibinfo{year}{2009}) \bibinfo{pages}{102--105}.
\bibitem[{Ostapchenko(2006)}]{qgsjetii}
\bibinfo{author}{S.~Ostapchenko},
\newblock \bibinfo{title}{{QGSJET-II}: towards reliable description of very
  high energy hadronic interactions},
\newblock \bibinfo{journal}{Nucl. Phys. B (Proc. Suppl.)} \bibinfo{volume}{151}
  (\bibinfo{year}{2006}) \bibinfo{pages}{143--146}.
\bibitem[{Aab et~al.(2014)}]{longXmax_fits2014}
  \bibinfo{author}{A.~Aab}, et~al., \newblock \bibinfo{title}{{Depth
      of maximum of air-shower profiles at the Pierre Auger
      Observatory. II. Composition implications}}, \newblock
  \bibinfo{journal}{Phys.Rev.}  \bibinfo{volume}{D90}
  (\bibinfo{year}{2014})
  \bibinfo{pages}{122006}. (\bibinfo{collaboration}{Pierre Auger
    Collaboration}).
\bibitem[{Riehn et~al.(2015)Riehn, Engel, Fedynitch, Gaisser, and
  Stanev}]{sibyll23_2015}
\bibinfo{author}{F.~Riehn}, et~al.,
\newblock \bibinfo{title}{{A new version of the event generator
    Sibyll, in: Proceedings of 34th International
  Cosmic Ray Conference, the Hague,}}
  \bibinfo{year}{2015}. \bibinfo{note}{arXiv:1510.00568}.
\bibitem[{Bergmann et~al.(2007)Bergmann, Engel, Heck et~al.}]{conex_2006}
\bibinfo{author}{T.~Bergmann}, \bibinfo{author}{R.~Engel},
  \bibinfo{author}{D.~Heck}, et~al.,
\newblock \bibinfo{title}{{One-dimensional hybrid approach to extensive air
  shower simulation}},
\newblock \bibinfo{journal}{Astropart. Phys.} \bibinfo{volume}{26}
  (\bibinfo{year}{2007}) \bibinfo{pages}{420--432}.
\bibitem[{Pierog et~al.(2006)Pierog, Alekseeva, Bergmann et~al.}]{conex_pylos}
\bibinfo{author}{T.~Pierog}, \bibinfo{author}{M.~K. Alekseeva},
  \bibinfo{author}{T.~Bergmann}, et~al.,
\newblock \bibinfo{title}{First results of fast one-dimensional hybrid
  simulation of {EAS} using {CONEX}},
\newblock \bibinfo{journal}{Nucl. Phys. B (Proc. Suppl.)} \bibinfo{volume}{151}
  (\bibinfo{year}{2006}) \bibinfo{pages}{159--162}.
\bibitem[{Abbasi et~al.(2015)}]{WGmass_UHECR2014}
\bibinfo{author}{R.~Abbasi}, et~al.,
\newblock \bibinfo{title}{{Report of the Working Group on the Composition of
  Ultra High Energy Cosmic Rays}}  (\bibinfo{year}{2015}). (\bibinfo{collaboration}{Pierre Auger
  and Telescope Array Collaborations}).
\bibitem[{Wibig and Sobczynska(1998)}]{wibig_cx1998}
\bibinfo{author}{T.~Wibig}, \bibinfo{author}{D.~Sobczynska},
\newblock \bibinfo{title}{{Proton-nucleus cross-section at high-energies}},
\newblock \bibinfo{journal}{J. Phys.} \bibinfo{volume}{G24}
  (\bibinfo{year}{1998}) \bibinfo{pages}{2037--2047}.
\bibitem[{Aab et~al.(2014{\natexlab{a}})}]{longXmax2014}
  \bibinfo{author}{A.~Aab}, et~al., \newblock \bibinfo{title}{{Depth of maximum of air-shower
      profiles at the Pierre Auger Observatory. I. Measurements at
      energies above $10^{17.8}$ eV}}, \newblock
  \bibinfo{journal}{Phys.Rev.} \bibinfo{volume}{D90}
  (\bibinfo{year}{2014}{\natexlab{a}}) \bibinfo{pages}{122005}.
  (\bibinfo{collaboration}{Pierre Auger Collaboration}).
\bibitem[{Aab et~al.(2014{\natexlab{b}})}]{mpd2014}
  \bibinfo{author}{A.~Aab}, et~al., \newblock \bibinfo{title}{{Muons in air showers at the
      Pierre Auger Observatory: Measurement of atmospheric production
      depth}}, \newblock \bibinfo{journal}{Phys.Rev.}
  \bibinfo{volume}{D90} (\bibinfo{year}{2014}{\natexlab{b}})
  \bibinfo{pages}{012012}.  (\bibinfo{collaboration}{Pierre Auger
    Collaboration}).
\bibitem[{Aab et~al.(2015)}]{has_muons2015} \bibinfo{author}{A.~Aab},
  et~al., \newblock \bibinfo{title}{{Muons in air showers at the
      Pierre Auger Observatory: Mean number in highly inclined
      events}}, \newblock \bibinfo{journal}{Phys.Rev.}
  \bibinfo{volume}{D91} (\bibinfo{year}{2015})
  \bibinfo{pages}{032003}.  (\bibinfo{collaboration}{Pierre Auger
    Collaboration}).
\bibitem[{Kampert et~al.(2013)}]{crpropa2}
\bibinfo{author}{K.-H. Kampert}, et~al.,
\newblock \bibinfo{title}{{CRPropa 2.0 -- a Public Framework for Propagating
  High Energy Nuclei, Secondary Gamma Rays and Neutrinos}},
\newblock \bibinfo{journal}{Astropart. Phys.} \bibinfo{volume}{42}
  (\bibinfo{year}{2013}) \bibinfo{pages}{41--51}.
\bibitem[{Puget et~al.(1976)Puget, Stecker, and Bredekamp}]{puget1976}
\bibinfo{author}{J.~L. Puget}, \bibinfo{author}{F.~W. Stecker},
  \bibinfo{author}{J.~H. Bredekamp},
\newblock \bibinfo{title}{{Photonuclear Interactions of Ultrahigh-Energy Cosmic
  Rays and their Astrophysical Consequences}},
\newblock \bibinfo{journal}{Astrophys. J.} \bibinfo{volume}{205}
  (\bibinfo{year}{1976}) \bibinfo{pages}{638--654}.
\end{thebibliography}
\end{document}